# Release of Brain Mitochondrial Hexokinase by Acidic Proteins and Macromolecular Polyanions

F. Moller

Department of Biochemistry, Queen=s University, Kingston, Ontario, Canada, K7L 3N6

E-mail: mollerf@post.queensu.ca.  Fax: (613) 533 2497.

**Abstract.** Preparations of arachidonic acid binding and non-binding proteins from bovine brain, four acidic proteins (α-casein, phosvitin, β-lactoglobulin A and B), the peptide polyglutamate, and two polyanions (heparin, dextran sulfate) enhanced both basal and glucose 6-phosphate induced solubilization of rat brain mitochondrial hexokinase (ATP:D-hexose 6-phosphotransferase, EC 2.7.1.1). In contrast, three other acidic proteins, had little (α-lactalbumin) or no effect (bovine serum albumin, ovalbumin) and five basic proteins inhibited release of the enzyme. Solubilizing activity therefore appears to require a net negative charge and one or more of the following structural features: extended conformation, random coil, unordered or β-structure, in this case the β-barrel in the fatty acid binding proteins and β-lactoglobulins. It is of interest that a difference of a single negative charge between β-lactoglobulin A and B, resulted in a statistically significant difference in the stimulation of hexokinase release. Possible physiological and pathological roles of this solubilizing effect are discussed briefly.

## Introduction

Hexokinase (HK) type 1 (e.g. Abrain hexokinase@) binds to the hexokinase binding protein (Felgner et al.1979), aka mitochondrial porin (Fiek et al.1982, Linden et al.1982, reviews by Wilson 1995, 2003), in the outer mitochondrial membrane. The enzyme can be solubilized from this site by, most importantly, G6P and ATP (Rose and Warms 1967, Wilson 1968, 1980). Since the soluble form is less active than the bound form (Wilson 1978, 1980) it has been suggested that the solubilization by G6P and the counteraction of this, particularly by inorganic phosphate (Rose and Warms 1967, Wilson 1968), may exert metabolic control of the enzyme in the brain and some other tissues. $Mg^{2+}$ at low, millimolar concentrations similarly inhibits solubilization by G6P possibly via screening of repulsive negative charges on the enzyme and the mitochondrion (Felgner and Wilson 1977, Wilson 1980). Type 2 HK (insulin- inducible hexokinase) also binds to the mitochondrion and behaves in a similar, although not identical, manner in insulin sensitive tissues (Wilson 1995, 2003). The importance of the above type of control depends, in part, clearly on the extent to which the enzyme can be solubilized by G6P and this has been found to vary in different species (Kabir and Wilson 1993, 1994). In the context of this paper it is of further interest that the soluble, but not the bound, form of the enzyme is susceptible to proteolytic breakdown (Crinelli et al. 1997, Polakis and Wilson 1985).

More recently an important mechanism of control of this enzyme has been shown to involve coupling of glucose phosphorylation by the bound hexokinase to oxidative phosphorylation (BeltrandelRio and Wilson 1991, 1992, De Cerquiera and Wilson 1995, 1998, Laterveer et al. 1995, 1997, Wicker et al.1993). There is evidence that this mechanism involves a mitochondrial porin as well as adenine nucleotide translocase (Beutner et al. 1996).

In view of the above it was of interest when it was found that preparations of rat liver and brain fatty acid binding proteins (FABPs) enhanced basal and G6P- stimulated release of hexokinase from rat brain mitochondria (see Results). The present paper deals with the solubilization of mitochondrial HK by some arachidonic acid binding proteins and some other acidic proteins and macromolecular polyanions.

## Materials and methods.

**Materials**. Rat liver and brain FABPs were prepared using brains from 300g male Sprague Dawley rats. Bovine brains were obtained from healthy heifers at slaughter. The bovine brains were placed in ice within 15 minutes of slaughter of the animals. Enzymes (snake venom phosphodiesterase (PDE, orthophosphoric diesterphosphohydrolase, EC 3.1.4.1), turkey egg white lysozyme (muramidase, mucopeptide N-acetyl- muramoylhydrolase, EC 3.2.1.17), micrococcal nuclease from Staphylococcus aureus (micrococcal endonuclease, EC 3.1.31.1), RNase (ribonuclease A, ribonucleate 3'-pyrimidinooligo-nucleotidohydrolase EC 3.1.27.5)) and other proteins, coenzymes, and Sephacryl S-100 HR were from Sigma Chemical Company, St. Louis, Missouri. Additional chromatography media were from Pharmacia Canada, Dorval, Quebec. Ultrafiltration membranes and Centricon 3 Microconcentrators were from Amicon Canada, Oakville, Ontario. Arachidonic acid [5,6,8,9,11,12,14,15-$^3$H(N)] was obtained from NEN Research Products, Boston, Massachusetts.

**Preparations.** All procedures were carried out at or below 5°C and preparations, at any stage, were stored at - 20°C (e.g. brain cytosol fractions) or in liquid nitrogen (minced grey matter). Ethylene glycol was added to some preparations to a concentration of 15 % (v/v). Highly purified rat liver and brain FABP were prepared by classical methods (Ockner et al. 1982).

Bovine brain cytosol fraction was obtained by homogenization of minced grey matter in three volumes of buffer (0.32 M sucrose, 5mM Tris-HCl, 1mM EGTA, 0.85 mM PMSF, pH 7.4) and ultracentrifugation. Using a PM-10 ultrafilter the cytosol fraction was concentrated and the concentrate separated on a short column of Sephacryl S-100 HR (referred to as Sephacryl Column1 in the following) with buffer AA@ (10 mM Tris-HCl, 0.15 M NaCl, pH 7.4). This separation produced the elution profile shown in Fig. 1A (see figure for further details). The fractions appearing in the cytochrome c position (the "shoulder" of the descending limb of the elution profile) were pooled ("Peak"A1 in the following) and, after concentration, and addition of arachidonic acid [5,6,8,9,11,12,14,15-$^3$H(N)], the concentrate was rechromatographed on a long column of Sephacryl S-100 HR (referred to as Sephacryl Column 2 below) using buffer AA@ (see Fig. 1B for details). The high molecular weight, radiolabelled material in "Peak"B1 from this column (Fig. 1B) was not further studied (it may contain aggregated fatty acid binding proteins (Fournier et al. 1983)). The radiolabelled fractions from the cytochrome c position on the other hand were combined (Peak B2, see Fig. 1B) and after concentration on an Amicon PM-10 ultrafilter, and exchange of the buffer with buffer AB@ (30 mM Tris-HCl, 15% ethylene glycol vol/vol, pH 8.5) placed on a DEAE Sepharose column. The DEAE column was next eluted with buffer AB@ followed by a 0-0.25M NaCl gradient in this buffer (see Fig. 2 for details). The fractions appearing in the cytidine position on Sephacryl Column 1 (combined in Peak A2, see Fig.1 A) were concentrated (Amicon YM-2 membrane) and rechromatographed on a long column of Sephacryl S-100 R (Sephacryl Column 3, Fig. 3) but without labelling with arachidonic acid. Only the higher molecular weight fractions of "Peak 1" from the latter column contained protein as determined by the Lowry assay (see Fig. 3 for details). All major peaks from the DEAE column, and some of the pooled fractions ("Peaks") from the three Sephacryl columns, were assayed for HK-releasing activity (see Table 1 below). **Assays**. The buffer in all preparations assayed for activity was exchanged with Hepes buffer (20mM Hepes, 15% ethylene glycol (vol/vol) pH 7.0) in order to eliminate effects of increased concentrations of salts on HK release (Felgner and Wilson 1977). Commercial protein/peptide preparations and heparin and dextran sulfate were dissolved in, and dialyzed against 10mM Hepes, pH 7.0 or transferred to this buffer using Centricon 3 microconcentrators. Rat brain mitochondria were prepared as described by Moller and Wilson (1983). Basal and G6P-stimulated release of the mitochondrial hexokinase were as described by Moller and Wilson (1983). Total HK activity (i.e. solubilized plus remaining bound, including latent activity released by 0.5% Triton X-100) was also determined. The average mitochondrial protein concentration in these assays, determined by the Biuret assay (Gornall et al. 1949), was $0.9 \pm 0.08$ mg/mL, n=13. Other protein determinations were carried out by the Lowry assay (Lowry et al. 1951). Electrophoresis under denaturing conditions was as described by Laemmli (1970). Student's t-test was used to determine significance of differences when possible.

## Results

Preliminary experiments showed that rat liver FABP (single band by electrophoresis) at 2.5 mg per mL doubled basal HK release and increased G6P-stimulated release 41% (n=5, $p<0.05$ and $p<0.001$ respectively by paired variate analysis). Furthermore rat brain FABP in one experiment tripled basal and doubled G6P-stimulated HK release. Two other smaller rat brain proteins, obtained in very small amounts, similarly increased HK release (results not shown).

The DEAE column chromatography (Column 3) of bovine brain preparations resulted in three major (i.e. "Peaks@ 4, 5,and 8), and several minor, arachidonic acid labelled peaks as well as three major unlabelled peaks ("Peaks" 2, 6, and 7, see Fig. 2 for details). It may be noted at this point that Schoentgen et al. (1989) found microheterogeneity in FABP purified from bovine brain. All these preparations caused release of HK whether they bound arachidonic acid or not thus suggesting a nonspecific, likely charge related, effect (Table 1). Both basal and G6P-stimulated release were enhanced by these preparations except when basal release appeared to be maximally stimulated as in the two cases of preparations from Peak 2, Column 1 (see Table 1). Hexokinase activity *per se* was also slightly increased in the presence of G6P by these particular preparations as shown by increases of total activity over controls (Table 1, and results not shown). In contrast, HK activity was not affected by the other bovine brain preparations as suggested by recoveries (total activities) close to 100% of controls (see Table 1). That fatty acid binding by some of these proteins was not involved in solubilizing the enzyme was suggested by the finding that treatment of rat liver FABP and protein in peak 2 from column 2 (see Fig.1 B) with Lipidex 1000 to remove bound fatty acids (Glatz and Veerkamp 1983) was without effect on their HK-releasing activity (results not shown). HK-releasing activity was independent of elution position on the DEAE column and thus apparently also of differences of (net negative) charge on the proteins.

**Hexokinase releasing activity of other proteins and some polyanions.** Since it was possible, given these data, that the HK-releasing activity was restricted to proteins with a net negative charge some polyanions and some proteins with either a net negative or positive charge at pH 7 were examined next for activity. Table 2 shows that the polyanions heparin and dextran sulfate, both with a high negative charge density, as well as the highly negatively charged peptide polyglutamate had high activity. Additionally, heparin and dextran sulphate also inhibited the hexokinase as shown by lower total activities (recoveries) than the controls. Presumably the sulfate groups were responsible for this inhibition since no inhibition was noted in the presence of compounds carrying only carboxylate groups or carboxylate and phosphate groups. The activity of some other (net) negatively charged, but less so, proteins (Table 3) varied from moderate (the phosphoproteins α-casein, and phosvitin, as well as the β-lactoglobulins) to low (α-lactalbumin) or virtually non- existent (BSA, and ovalbumin).

In contrast all five positively charged (at pH 7) proteins tested, except apparently snake venom phosphodiesterase, inhibited basal HK release and all five, without exception, inhibited release in the presence of G6P (Table 4). The apparent increase of basal release by the phosphodiesterase (Table 4) was most likely due to lysis of the mitochondria with concomitant solubilization of proteins, including hexokinase. All five positively charged proteins caused agglutination of the mitochondria, but only the phosphodiesterase caused lysis (results not shown).

## Discussion

Since not all the negatively charged proteins tested increased the release of the hexokinase and since positively charged proteins were inhibitory in this regard (possibly due to occlusion of the agglutinated mitochondria, however) it is concluded that a net negative charge is a necessary, but not sufficient, condition for HK-releasing activity of a protein. Thus certain structural features appear also to be required. Based on the structures of the proteins tested these would include: A random coil structure in aqueous solution as in polyglutamate (Iizuka and Yang 1965), open flexible chains especially in phosphate carrying regions as in casein and phosvitin (Holt and Sawyer 1988, Holt 1992)), and β-structure, especially the β-barrel as in the β-lactoglobulins (Papiz et al. 1986, Banaszak et al. 1994). The α-casein (a mixture of 60% $α_{s1} + α_{s0}$ casein and 40% β-casein with 8-9 and 5 mol P/ mol repectively arranged in clusters of two or four (Holt 1992)) had moderately high activity similar to polyglutamate. Considering this the low activity of phosvitin which has nearly enough phosphate groups to give a negative charge density of one per amino acid residue (Taborsky 1983) was a surprise. It is possible that this low activity is due to aggregate formation of the protein during frozen storage (Taborsky 1974). Of great interest, and needing further work, is the finding that the A form of β-lactoglobulin, which has just one more negative charge (in the β-barrel) than the B form (Braunitzer et al. 1972), also is the most active of the two (Table 2). Significantly, the β-barrel is also a prominent structural feature of the FABPs (Banaszak et al.1994). This circumstance and the finding that the protein must have a net negative charge add to the evidence that the FABPs indeed have this activity. It would be of interest to study the effect of phosphorylation on the activity of these proteins. The lack of activity of BSA may be due to the α-helical and overall rigid tertiary structure, due to its many disulfide bonds, of this negatively charged (at pH 7) protein (Brown 1976, Carter and Ho 1994). The virtual lack of activity of ovalbumin is at odds with the criteria outlined above. This protein has a high negative charge (see data of Nisbet et al. 1981) including two phosphate groups (Vogel and Bridger 1982) and more than fifty percent β-structure (Stein et al. 1991). The lack of a β-barrel or, possibly, a diffuse distribution of its negative charges may perhaps explain its lack of activity. The (low) activity of α-lactalbumin may be due to its negative charge and fifty to sixty percent unordered structure (B variant, McKenzie and White 1991). Finally, both a high negative charge density and extended conformation are characteristics of glycosaminoglycans thus explaining the high activity of heparin and in a similar way of dextran sulfate.

**Possible applications of the findings**. The low concentration of FABPs in the brain (Bass et al. 1984, Bass and Manning 1986) probably precludes any role in the control of the distribution and activity of HK in brain tissue. However, such control may be possible e.g. in the heart and in skeletal muscle which have both a (constitutive) type 1 HK (Abrain HK@) and the insulin inducible type 2 HK ( Wilson 1995, 2003). As noted in the introduction both of these hexokinase isozymes bind to the mitochondrion and are released by G6P. By causing complete solubilization or perhaps rather a switch from firm binding to porin to a more loose binding to the enzyme via $Mg^{2+}$- Abridges@ FABPs could play a role in changing from glucose to fatty acid oxidation in heart and muscle and some other tissues which have both isozymes. Such a role would dovetail well with the function of FABPs as delivery vehicles for fatty acids to the mitochondrion (Fournier et al. 1978, Fournier and Rahim 1985).

Given that a net negative charge and certain structural features may impart HK-releasing activity to proteins and glycosaminoglycans (GAGs) pathological conditions may occur in the brain in which this activity may come into play. This may for example be the case with certain sulfated glycosaminoglycans (GAGs). Thus heparan sulfate (Snow et al. 1989, Goedert et al. 1996) and other sulfated GAGs (Bonay and Avila 2001) may be present intraneuronally in the brain in Alzheimer=s Disease (AD). It is of interest in this connection that various sulfated glycosaminoglycans (Goedert et al. 1996 and Hasegawa et al. 1997) and polyanions like dextran sulfate and RNA (Kampers et al.1996 and Hasegawa et al. 1997) stimulate hyperphosphorylation and aggregation of tau proteins *in vitro*. Furthermore, since a single addition of a negative charge may increase the HK-releasing effect of a protein it is likely that addition of several negative charges as in hyperphosphorylation would be even more effective in this respect. Thus hyperphosphorylated, acidic, rodlike tau proteins (Cleveland et al. 1977) have been found intracellularly in AD (Friedhoff et al. 2000, Köpke et al. 1993, Sato et al. 2001) and other dementias with tau pathology (Goedert and Spillantini 2000, Delacourte and Buée 2000). As a result solubilization of the hexokinase could occur rendering the enzyme both less active (Wilson 1978, 1980) and subject to proteolytic breakdown (Polakis and Wilson 1985, Crinelli et al.1997). In addition as indicated in the introduction, and probably more important, coupling to oxidative phosphorylation could be interrupted causing lack of control of glucose oxidation. Supporting these possibilities are findings that glucose metabolism is decreased in AD (Piert et al. 1996) even when correction is made for atrophy of the brain (Alavi et al. 1993, Tanna et al.1991, Ibanez et al. 1998). Further support comes from findings that aggregated tau proteins besides being enzymatically glycosylated are highly glycated, i.e. nonenzymatically glycosylated (Ledesma et al. 1994, Ledesma et al. 1995, Nacharaju et al. 1997) suggesting exposure to sustained high levels of glucose. Considering that the brain depends virtually completely on glucose for its energy requirements a decreased rate of its metabolism may thus be a contributing factor in the progression of AD and other dementias involving accumulation of hyperphosphylated proteins such as the tau-proteins. Recent studies have demonstrated an antiapoptotic effect of mitochondrially bound hexokinases, i.e. either type 1 or type 2 (review: Pastorino and Hoek 2003, Azoulay-Zohar et al. 2004). Release of the enzymes by the charge- related mechanism outlined above could therefore lead to death of neurons in diseases such as AD and, in general, to cell death in other tissues than the brain.

**Acknowledgements**

The work in this report was supported by grants from the Medical Research Council of Canada (MA 9210) and from Queen's University. The author is especially grateful to his colleagues Dr. D.R. Forsdyke and Dr. D.J. Walton for making laboratory space available for parts of this research and for valuable advice and friendly support.

**References**

Alavi, A., Newberg, A.B., Souder, E., and Berlin, J.A. 1993. Quantitave analysis of PET and MRI data in normal aging and Alzheimer=s disease: atrophy weighted total brain metabolism and absolute whole brain metabolism as reliable discriminators. J. Nucl. Med. 34: 1681-1687.

Azulay-Zohar, H., Israelson,A., Abu-Hamad, S., and Shoshan-Barmatz, V. 2004. In self-defence: hexokinase promotes voltage-dependent anion channel closure and prevents mitochondria-mediated apoptotic cell death. Biochem. J. 347 - 355.

Banaszak, L., Winter. N., Xu, Z., Bernlohr, D.A., Cowan, S., and Jones, A. 1994. Lipid-binding proteins: A family of fatty acid and retinoid transport proteins. Adv. Prot. Chem. 45: 89 151.

Bass, N.M., and Manning, J.A. 1986. Tissue expression of three structurally different fatty acid binding proteins from rat heart, liver, and intestine. Biochem. Biophys. Res. Com. 137: 929-935.

Bass, N.M., Raghupathy, E., Rhoads, D.E., Manning, J.A., and Ockner, R.K. 1984. Partial purification of molecular weight 12000 fatty acid binding proteins from rat brain and their effect on synaptosomal $Na^+$-dependent amino acid uptake. Biochemistry 23: 6539 - 6544.

BeltrandelRio, H., and Wilson, J.E. 1991. Hexokinase of rat brain mitochondria: Relative importance of adenylate kinase and oxidative phosphorylation as sources of substrate ATP, and interaction with intramitochondrial compartments of ATP and ADP. Arch. Biochem. Biophys. 286: 183- 194.

BeltrandelRio, H., and Wilson, J.E. 1992. Coordinated regulation of cerebral glycolytic and oxidative metabolism, mediated by mitochondrially bound hexokinase dependent on intramitochondrially generated ATP. Arch. Biochem. Biophys. 296: 667-677.

Beutner, G., Rück, A., Riede, B., Welte, W., and Brdiczka, D. 1996. Complexes between kinases, mitochondrial porin and adenylate translocator in rat brain resemble the permeability transition pore. FEBS Lett. 396: 189-195.

Bonay, P., and Avila, J. 2001. Apolipoprotein E4 stimulates sulfation of glycosaminoglycans in neural cells. Biochim. Biophys. Acta 1535: 217-220,

Braunitzer, G., Chen, R., Schrank, B., and Stangl, A. 1972. Automatische Squenzanalyse eines Proteins (β-Lactoglobulin AB). Hoppe-Seyler=s Z. Physiol. Chem. 353: 832-834.

Brown, J.R. 1976. Structural origins of mammalian albumin. Fed. Proc. 35: 2141 - 2144.

Carter, D.C., and Ho, J.X. 1994. Structure of serum albumin. Adv Prot Chem 45: 153-203.

Cleveland, D.W., Hwo, S-Y., and Kirschner, M.W. 1977. Physical and chemical properties of purified tau factor and the role of tau in microtubule assembly. J Mol Biol 116: 227-247.

Colton, T. 1974. In: Statistics in Medicine. Little, Brown and Company. Boston. pp 39-40.

Crinelli, R., Bianchi, M., Serafini, G., and Magnani, M. 1997 Structural determinants that make hexokinase susceptible to ubiquitin- and ATP-dependent proteolysis. Biochem. Soc. Trans. 25: 69S.

De Cerquiera, C.M., and Wilson, J.E. 1995. Application of a double isotopic labeling method to a study of the interaction of mitochondrially bound rat brain hexokinase with intramitochondrial compartments of ATP generated by oxidative phosphorylation. Arch. Biochem. Biophys. 324: 9-14.

De De Cerquiera, C.M. and Wilson, J.E. 1998. Further studies on the coupling of mitochondrially bound hexokinase to intramitochondrially compartmented ATP generated by oxidative phosphorylation. Arch. Biochem. Biophys. 350: 109-117.

Delacourte, A., and Buée, L. 2000. Tau pathology: a marker of neurodegenerative disorders. Curr. Opin. Neurol. 13: 371-376.

Dolapchiev, L.B., Vassileva, R.A., and Koumanov, K.S. 1980. Venom exonuclease II. Amino acid composition and carbohydrate, metal ion and lipid content in the Crotalus adamenteus venom exonuclease. Biochim. Biophys. Acta 622: 331-336.

Felgner, P.L., and Wilson, J.E. 1977. Effect of neutral salts on the interaction of rat brain hexokinase with the outer mitochondrial membrane. Arch. Biochem. Biophys. 182: 282-294.

Felgner, P.L., Messer, J.L., and Wilson, J.E. 1979. Purification of a hexokinase-binding protein from the outer mitochondrial membrane. J. Biol. Chem. 254: 4946-4949.

Fiek, C., Benz, R., Roos, N., and Brdisczka, D. 1982. Evidence for identity between the hexokinase-binding protein and the mitochondrial porin in the outer membrane of rat liver mitochondria. Biochim. Biophys. Acta 688: 429-440.

Fournier, N.C., Geoffroy, M., and Deshusses, J. 1978. Purification and characterization of a long chain, fatty-acid-binding protein supplying the mitochondrial β-oxidative system in the heart. Biochim. Biophys. Acta 533: 457 - 464.

Fournier, N.C., and Rahim, M. 1985. Control of energy production in the heart: A new function for fatty acid binding protein. Biochemistry 24: 2387-2396.

Fournier, N.C., Zuker, M., Williams, R.E., and Smith, I.C.P. 1983. Self-association of the cardiac fatty acid binding protein. Influence on membrane-bound, fatty acid dependent enzymes. Biochemistry 22: 1863-1872.

Friedhoff, P., von Bergen, M., Mandelkow, E-M., and Mandelkow, E. 2000. Structure of tau protein and assembly into paired helical filaments. Biochim. Biophys. Acta 1502: 122-132.

Glatz, J.F.C., and Veerkamp, J.H. 1983. Removal of fatty acids from serum albumin by Lipidex 1000 chromatography. J. Biochem. Biophys. Methods 8: 57-61.

Goedert, M., Jakes, R., Spillantini, M.G., Hasegawa, M., Smith, M.J., and Crowther, R.A. 1996. Assembly of microtubule-associated protein tau into Alzheimer-like filaments induced by sulphated glycosaminoglycans. Nature 338: 550-553.

Goedert, M., and Spillantini, M.G. 2000. Tau mutations in frontotemporal dementia FTDP-17 and their relevance for Alzheimer=s disease. Biochim. Biophys. Acta 1502: 110-121.

Gornall, A.G., Bardawill, C.S., and David, M.M. 1949. Determination of serum proteins by means of the biuret reaction. J. Biol. Chem. 177: 751-766.

Hasegawa, M., Crowther, R.A., Jakes, R., and Goedert, M. 1997. Alzheimer-like changes in microtubule-associated protein tau induced by sulfated glycosaminoglycans. J. Biol. Chem. 272: 33118-33124.

Holt, C. 1992. Structure and stability of bovine casein micelles. Adv. Prot. Chem. 43: 63 - 151.

Holt, C., and Sawyer, L. 1988. Primary and predicted secondary structures of the caseins in relation to their biological functions. Prot. Eng. 2: 251-259.

Ibanez, V., Pietrini, P., Alexander, G.E., Furey, M.L., Teichberg, D., Rajapakse, J.C., Rapoport,

Iizuka, E., and Yang, J,T. 1965. Effect of salts and dioxane on the coiled conformation of poly-L-glutamic acid in aqueous solution. Biochemistry 4: 1249 - 1257.

Kabir, F., and Wilson, J.E. 1993. Mitochondrial hexokinase in brain of various species: differences in sensitivity to solubilization by glucose 6-phosphate. Arch. Biochem. Biophys. 300 (2): 641-650.

Kabir. F., and Wilson, J.E. 1994. Mitochondrial hexokinase in brain: Coexistence of forms differing in sensitivity to solubilization by glucose 6-phosphate on the same mitochondria. Arch. Biochem. Biophys. 310: 410-416.

Kampers, T., Friedhoff, P., Biernat, J., Mandelkow, E-M., and Mandelkow, E. 1996. RNA stimulates aggregation of microtubule-associated protein tau into Alzheimer-like paired helical filaments. FEBS Lett. 399: 344-349.

Kimelberg, H.K., Papahadjopoulos, D. 1971. Interactions of basic proteins with phospholipid membranes. J. Biol. Chem. 246: 1142-1148.

Köpke, E., Tung, Y., Shaikh, S., Alonso, A. delC., Iqbal, K., and Grundke-Iqbal, I. 1993. Microtubule-associated protein tau. Abnormal phosphorylation of a non-paired helical filament pool in Alzheimer=s disease. J. Biol. Chem. 268: 24374-24384.

Laemmli, U.K. 1970. Cleavage of structural proteins during the assembly of the head of bacteriophage T4. Nature 227: 680-685.

Laterveer, F,D., Gellerich, F.N., and Nicolay, K.  1995.  Macromolecules increase the channeling of ADP from externally associated hexokinase to the matrix of mitochondria. Eur. J. Biochem. 232: 569- 577.

Laterveer, F.D.,  Nicolay, K.,and Gellerich, F.N.  1997.  Experimental evidence for dynamic compartmentation of ADP at the mitochondrial periphery: Coupling of mitochondrial adenylate kinase with oxidative phosphorylation under conditions mimicking the intracellular osmotic pressure.  Mol. Cell. Biochem. 174: 43-51.

Ledesma, M.D., Bonay, P., Colaco, C., and Avila, J.  1994.   Analysis of microtubule-associated protein tau glycation in paired helical filaments.  J. Biol. Chem. 269: 21614-21619.

Ledesma, M.D., Bonay, P., and Avila, J.  1995.  τ protein from Alzheimer=s disease patients is glycated at its tubulin binding domain. J. Neurochem. 65: 1658-1664.

Linden, M., Gellerfors, P., and Nelson, B.D.  1982.  Pore protein and the hexokinase binding protein from the outer membrane of rat liver mitochondria are identical.  FEBS Lett. 141: 189-192.

Lowry, O.H., Rosebrough, N.J., Farr, A.L., and Randall, R.J.  1951.   Protein measurement with Folin phenol reagent.  J. Biol. Chem. 193: 265-275.

McKenzie, H.A., and White, F.H. Jr.  1991.  Lysozyme and α-lactalbumin: Structure, function, and interrelationships.  Adv.  Prot. Chem. 41: 173-315.

Moller, F., and Wilson, J.E.  1983.  The influence of specific phospholipids on the interaction of hexokinase with the outer mitochondrial membrane.  J. Neurochem. 41: 1109-1118.

Nacharaju, P., Ko, L-W., and Yen, S-H. C.  1997.  Characterization of in vitro glycation sites of tau.  J. Neurochem. 69: 1709-1719.

Nisbet, A.D., Saundry, R.H., Moir, A.J.G., Fothergill, L.A., and Fothergill, J.E. 1981. The complete amino-acid sequence of hen ovalbumin. Eur. J. Biochem. 115: 335-345.

Ockner, R.K., Manning, J.A., and Kane, J.P. 1982. Fatty acid binding protein. Isolation from rat liver, characterization, and immunological quantification. J. Biol. Chem. 257: 7872-7878.

Papiz, M.Z., Sawyer, L., Eliopoulos, E.E., North, A.C.T., Findley, J.B.C., Sivaprasadarao, R., Jones, T.A., Newcomer, M.E., and Kraulis, P.J. 1986. The structure of β-lactoglobulin and its similarity to plasma retinol-binding protein. Nature 324: 383-385.

Pastorino, J.G., and Hoek, J.B. 2003. Hexokinase II: The integration of energy metabolism and control of apoptosis. Curr. Med. Chem. 16: 1535 - 51.

Piert, M., Koeppe. R.A., Giordani, B., Berent, S., and Kuhl. D.E. 1996. Diminished glucose transport and phosphorylation in Alzheimer=s disease determined by dynamic FDG-PET. J. Nucl. Med. 37: 201-208.

Polakis, P.G., and Wilson, J.E. 1985. An intact N-terminal sequence is critical for binding of rat brain hexokinase to mitochondria. Arch. Biochem. Biophys. 236: 328-337.

Rose, I.A., and Warms, J.V.B. 1967. Mitochondrial hexokinase. Release, rebinding, and location. J. Biol. Chem. 242: 1635 - 1645.

Sato, Y., Naito,Y., Grundke-Iqbal, I., Iqbal, K., and Endo, T. 2001. Analysis of N-glycans of pathological tau: possible occurrence of aberrant processing of tau in Alzheimer=s disease. FEBS Lett. 496: 152-160. 2001.

Schoentgen, F., Pignède, D., Bonanno, L.M., and Jollès, P. 1989. Fatty acid binding protein from bovine brain. Amino acid sequence and some properties. Eur. J. Biochem. 185: 35-40.

Snow, A.D., Lara, S., Nochlin, D., and Wight, T.N. 1989. Cationic dyes reveal proteoglycans

structurally integrated within the characteristic lesions of Alzheimer=s Disease.  Acta Neuropathol. 78: 113-123.

Stein, P.E., Leslie, A.G.W., Finch, G.T., and Carrell, R.W.  1991.  Crystal structure of uncleaved ovalbumin at 1.95 Å resolution.  J. Mol. Biol. 221:  941 - 959.

Taborsky, G.  1974.  Phosphoproteins.  Adv. Prot. Chem. 28: 1- 210.

Taborsky, G.  1983.  Phosvitin.  Adv. Inorganic Biochem. 5: 235-279.

Taniuchi, H., Anfinsen, C.B., and Sodja, A.  1967.  The amino acid sequence of an extracellular nuclease of Staphylococcus aureus.  J. Biol. Chem. 242: 4752-4758.

Tanna, N.K., Kohn, M.I., Horvich, D.N., Jolles, P.R., Zimmerman, R.A., Alves, W.M., and Alavi, A.  1991.  Analysis of brain and cerebrospinal fluid volumes with MR imaging : impact on PET data correction for atrophy, II: aging and Alzheimer=s disease.  Radiology 178: 123-130.

Vogel, H.J., and Bridger, W.A.  1982.  Phosphorus-31 nuclear magnetic resonance studies of the two phosphoserine residues of hen egg white albumin.  Biochemistry 21: 5825 - 5831.

Wicker, U., Bucheler, K., Gellerich, F.N., Wagner, M., Kapischke, M., and Brdiczka, D.  1993.  Effect of macromolecules on the structure of the mitochondrial inter-membrane space and the regulation of hexokinase.  Biochim. Biophys. Acta 1142: 228-239.

Wille, H., Drewes, G., Biernat, J., Mandelkow, E.M., and Mandelkow, E.  1992.  Alzheimer-like paired helical filaments and antiparallel dimers formed from microtubule-associated protein tau in vitro.  J. Cell Biol. 118: 573 - 584.

Wilson, J.E.  1968.  Brain Hexokinase. A proposed relation between soluble-particulate distribution <u>in vivo</u>.  J. Biol. Chem. 243: 3640-3647.

Wilson, J.E. 1978. Ambiquitous enzymes: Variation in intracellular distribution as a regulatory mechanism. Trends Biochem. Sci. 3: 124-125.

Wilson, J.E. 1980. Brain hexokinase, the prototype ambiquitous enzyme. Curr. Top. Cell. Regul. 16: 1 - 44.

Wilson, J.E. 1995. Hexokinases. Rev. Physiol. Biochem. Pharmacol. 126: 65-198.

Wilson, J.E. 2003. Isozymes of mammalian hexokinase: Structure, subcellular localization, and metabolic function. J. Exp. Biol. 206: 2049 - 57.

**Table 1.** Solubilization of rat brain mitochondrial hexokinase activity by bovine brain preparations.

| Column / Peak | Protein mg/mL | Released Hexokinase % of Control | | Total Hexokinase % of Control | |
|---|---|---|---|---|---|
| | | -G6P | +G6P | -G6P | +G6P |
| 1 / 2 | 3.0 | 1325 | 84  | 103 | 118 |
| 1 / 2 | 3.0 | 1416 | 114 | 107 | 113 |
| 2 / 2 | 1.5 | 186  | 149 | 100 | 94  |
| 3 / 1 | 1.5 | 155  | 141 | 100 | 98  |
| 3 / 2 | 1.5 | 183  | 144 | 96  | 100 |
| 3 / 3 | 1.5 | 184  | 138 | 97  | 103 |
| 3 / 4 | 1.5 | 396  | 220 | 98  | 93  |
| 3 / 5 | 1.5 | 361  | 193 | 96  | 97  |
| 3 / 6 | 1.5 | 284  | 178 | 104 | 104 |
| 3 / 7 | 1.5 | 225  | 211 | 96  | 102 |
| 3 / 7 | 3.0 | 340  | 272 | 96  | 108 |
| 3 / 8 | 1.5 | 209  | 203 | 97  | 109 |
| 3 / 8 | 3.0 | 324  | 230 | 95  | 106 |
| 4 / 1 | 1.5 | 491  | 310 | -   | -   |
| 4 / 1 | 1.5 | -    | 365 | -   | -   |

The results are from eight different experiments. Mean hexokinase activities (units per mL$\times 10^{-4}$ ± SD) in the controls, i.e. without added brain preparation, were: Basal, 0.17±0.07; G6P-stimulated, 1.94± 0.48; total, 7.82±0.67. Precision of duplicate determinations (Colton 1974) using the results in Table 1 and measured by the average standard deviation was ± 8.3 × $10^{-6}$ units/mL in the absence -and ± 24.2 × $10^{-6}$ units/mL in the presence of G6P.

**Table 2.** Solubilization of rat brain mitochondrial hexokinase by three negatively charged polymers.

| Compound / n | Conc. mg/mL | Released Hexokinase % of Control | | Total Hexokinase % of Control | |
|---|---|---|---|---|---|
| | | −G6P | +G6P | −G6P | +G6P |
| Heparin/4 | 3 | 2086 $p<0.01$ | 213 $p<0.001$ | 78 | 83 |
| Dex. Sulfate/3 | 3 | 1856 $p<0.01$ | 213 $p<0.05$ | 69 | 78 |
| Polyglutam./3 | 3 | 512 $p<0.02$ | 260 $p<0.02$ | 100 | 105 |

The results are from 10 different experiments. Average released activities (units per mL $\times 10^{-4} \pm$ SD) of controls were: Basal, 0.18 ±0.03; G6P-stimulated, 2.03 ± 0.24; total, 6.76 ±0.50. The number of duplicate determinations is indicated by n, and p is the probability of the difference between a results and its control being zero. See Methods for further details.

**Table 3**. Solubilization of rat brain mitochondrial hexokinase by some acidic proteins (pI<7)[a]

| Compound /n | Conc. mg/mL | Released Hexokinase % of Control | | Total Hexokinase % of Control | |
|---|---|---|---|---|---|
| | | −G6P | +G6P | −G6P | +G6P |
| BSA /1 | 1 | 97 | 90 | 112 | 103 |
| BSA /1 | 3.75 | 90 | 102 | nd | nd |
| BSA /1 | 15 | 100 | 103 | nd | nd |
| α-casein /3 | 3 | 430 $p<0.01$ | 259 $p<0.001$ | 94 | 113 |
| α-lactalbumin /2 | 3 | 132 p: nd | 135 p: nd | 98 | 101 |
| β-lactoglob. A /3 | 3 | 307 $p<0.02$ | 173 $p<0.01$ | 97 | 110 |
| β-lactoglob. B /3 | 3 | 197 $p<0.01$ | 136 $p<0.01$ | 100 | 107 |
| phosvitin /3 | 3 | 300 $p<0.02$ | 173 $p<0.02$ | 106 | 105 |
| ovalbumin /2 | 3 | 96 p: nd | 121 p: nd | 97 | 102 |

[a] The difference between the stimulation of basal release by β-lactoglobulin A and B was significant at the 0.05 level and in the presence of of G6P at the 0.01 level. See footnote to Table 2 and Methods for further details.

**Table 4**. Effect of some basic proteins on release of mitochondrial hexokinase.

| Protein | Conc. mg/mL | Released Hexokinase % Control Activity | | Total Hexokinase % Control Activity | | pI [Ref.] |
|---|---|---|---|---|---|---|
| | | - G6P | + G6P | - G6P | + G6P | |
| Cytochrome c | 0.50 | 112 | 62 | 115 | 108 | 10.7[a] |
| | 1.00 | 44 | 39 | 105 | 101 | |
| Lysozyme | 0.25 | 61 | 50 | 105 | 102 | 11.0[a] |
| | 0.50 | 32 | 23 | 89 | 87 | |
| RNase | 0.50 | 98 | 73 | 106 | 105 | 9.7[a] |
| | 1.00 | 52 | 44 | 99 | 101 | |
| Phospho-diesterase | 0.50 | 220 | 77 | 119 | 116 | 9.0[b] |
| | 1.00 | 142 | 59 | 117 | 114 | |
| Micrococcal Endonuclease | 0.25 | 80 | 50 | 101 | 101 | 10.4[c] |
| | 0.50 | 63 | 38 | 103 | 97 | |

Average activities released (units per mL x $10^{-4}$) of controls were: basal, 0.325; G6P-stimulated, 2.58; total, 10.83. Isoelectric points cited in: [a] Kimelberg and Papahadjopoulos (1971); [b] Dolapchiev et al. 1980; [c] calculated from amino acid sequence in: Taniuchi et al. 1967. See text for further details

**Fig. 1**

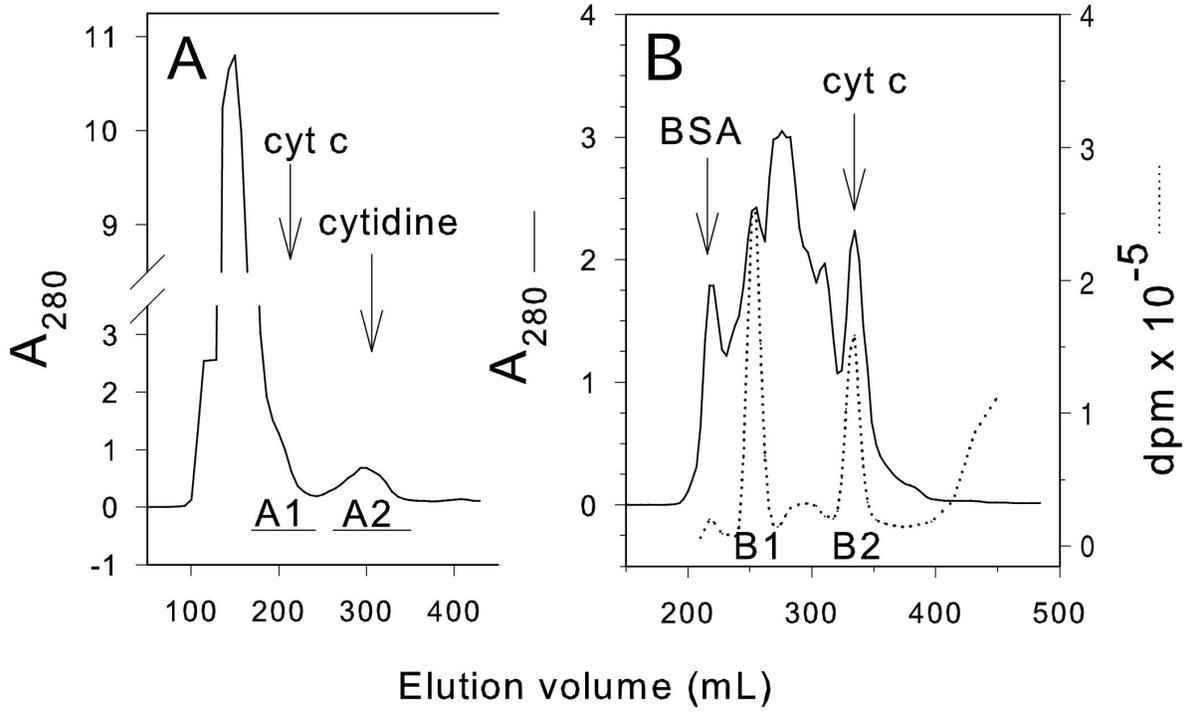

**Fig. 2**

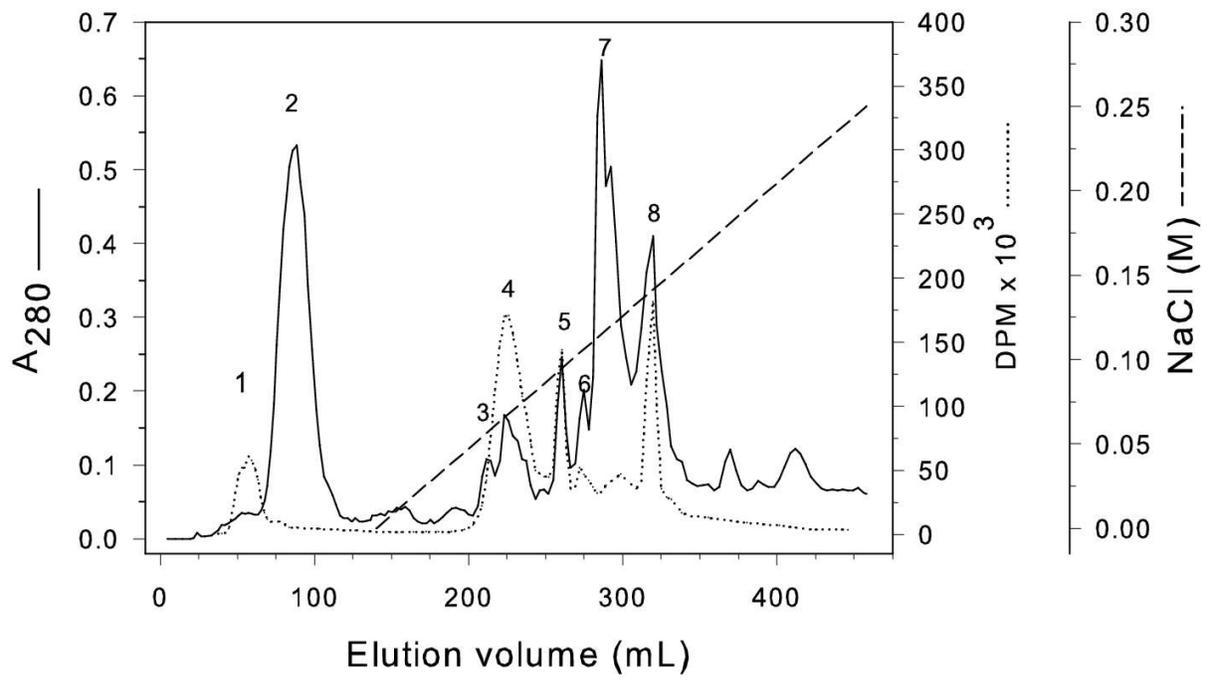

**Fig. 3**

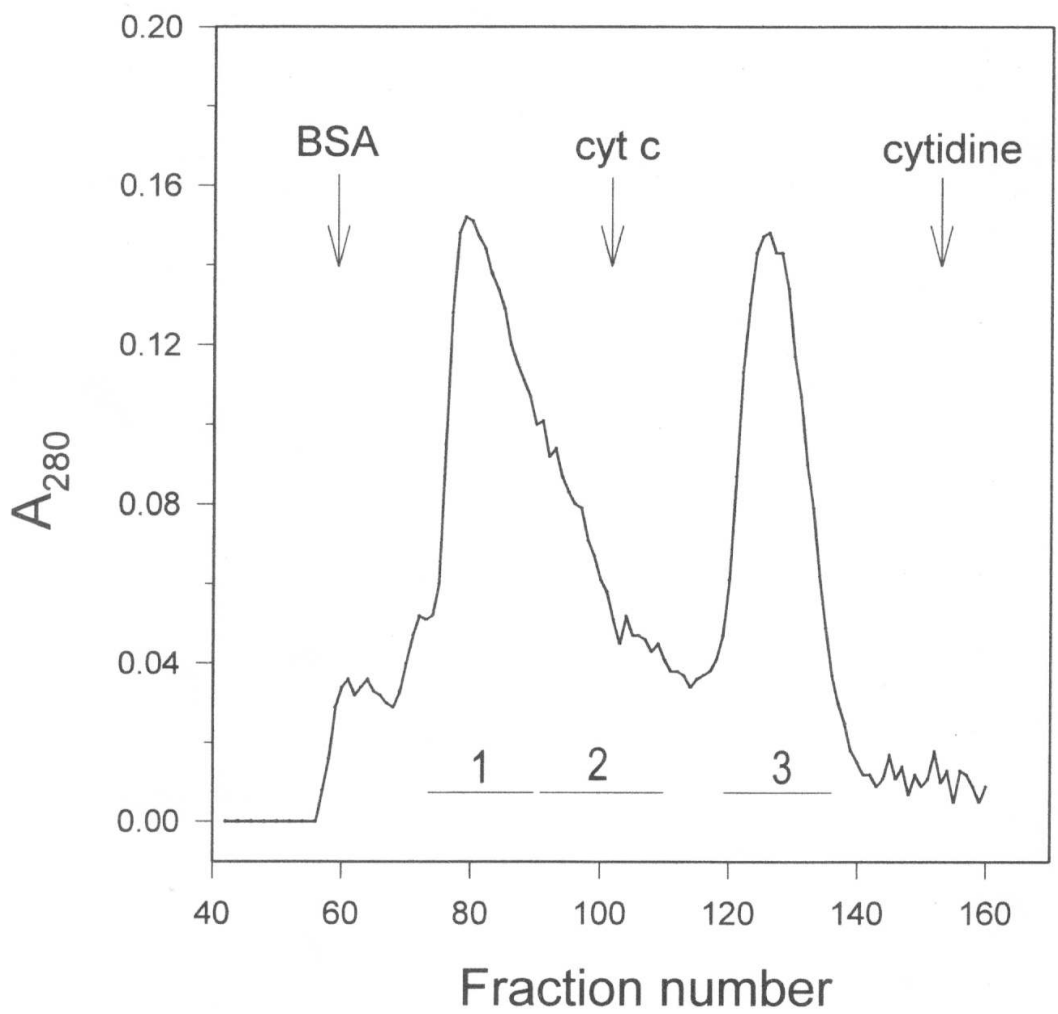

# Text to Figures

**Fig. 1.** Fractionation of bovine brain cytosol fraction. **A**. **Sephacryl Column 1**. Concentrated cytosol fraction (684 mg protein in 11 mL) was applied to, and eluted on, a 5cm ⌀ x18 cm column of Sephacryl S-100 HR with buffer A (see Methods). The column was calibrated with a mixture of Blue Dextran, BSA (not shown), cytochrome c, and cytidine. **"Peaks" 1** and **2** correspond to number of fractions of 7.1 mL indicated by the bars in the figure. **B**. **Sephacryl Column 2**. To the concentrated "Peak" 1 (10mL, 437 mg protein), i.e. shoulder at cytochrome c position, from column 1 was added tritium-labeled arachidonic acid (1 µCi in 10 µL ethanol at 0° C) and the mixture eluted on a 2.5cm ⌀ x 95cm column of Sephacryl S-100 HR with buffer A. The column was calibrated as above. **"Peaks" 1** and **2** correspond to number of fractions (3.25 mL) pooled as indicated by the bars.

**Fig. 2. Column 3.** Concentrated Peak 2, i.e. from cytochrome c position on column 2 (49 mg protein in 16 mL), was transferred to buffer B and applied slowly to a column (1.6cm ⌀ x 30cm) of DEAE Sepharose CL6B equilibrated with the same buffer (see Methods). The column was eluted with two bed volumes buffer B at which point a NaCl gradient (0-0.25M) was superimposed and five more bed volumes collected. Fractions of 2.8 mL were pooled (**Peaks 1, 2,** etc.) corresponding to the numbers above the peaks.

**Fig. 3. Column 4.**(Sephacryl Column 3) Fractionation of "Peak 2" from Column 1 on Sephacryl S-100 HR: 2.5 mg protein in 0.4mL, was applied to the column ( 0.92cm ⌀ x 125cm) equilibrated and then eluted with buffer B. Fractions of 0.56 mL were pooled (**Peaks 1, 2,** and **3**) as indicated by the bars.